\documentclass[journal,twoside,web]{ieeecolor}
\usepackage{generic}
\usepackage{cite}
\usepackage{amsmath,amssymb,amsfonts}
\usepackage{algorithmic}
\usepackage{graphicx}
\usepackage{textcomp}
\usepackage{bm}  
\usepackage{array}
\usepackage{graphicx}
\usepackage{algorithm}
\usepackage{setspace}
\usepackage{multirow}
\usepackage{dcolumn}
\usepackage{booktabs}
\usepackage{color}

\def\BibTeX{{\rm B\kern-.05em{\sc i\kern-.025em b}\kern-.08em
    T\kern-.1667em\lower.7ex\hbox{E}\kern-.125emX}}
\begin{document}
\title{One-step Method for Material Quantitation using In-line Tomography with Single Scanning}
\author{Suyu Liao, Shiwo Deng, Yining Zhu,  Huitao Zhang, Peiping Zhu, Kai Zhang, and Xing Zhao
\thanks{This work was supported in part by National Natural Science Foundation of China (No. 61971293, 61827809 and 61671311), the Sino-German Center (M-0187) and National Key Research and Development Program of China (No. 2020YFA0712200).(Corresponding author: Yining Zhu and Huitao Zhang.) }
\thanks{ Suyu Liao is with School of Mathematical Sciences, Capital Normal University, Beijing, 100048, China. }
\thanks{ Shiwo Deng is with School of Mathematical Sciences, Capital Normal University, Beijing, 100048, China and Shenzhen National Applied Mathematics Center, Southern University of Science and Technology, Shenzhen, China. }
\thanks{ Yining Zhu is with School of Mathematical Sciences, Capital Normal University, Beijing, 100048, China and Shenzhen National Applied Mathematics Center, Southern University of Science and Technology, Shenzhen, China.(e-mail: ynzhu@cnu.edu.cn) }
\thanks{ Huitao Zhang is with School of Mathematical Sciences, Capital Normal University, Beijing, 100048, China and Shenzhen National Applied Mathematics Center, Southern University of Science and Technology, Shenzhen, China.(e-mail: zhanght@cnu.edu.cn) }
\thanks{ Peiping Zhu and Kai Zhang is with Institute of High Energy Physics, Chinese Academy of Sciences, Beijing, 100049, China. }
\thanks{ Xing Zhao is with School of Mathematical Sciences, Capital Normal University, Beijing, 100048, China and Shenzhen National Applied Mathematics Center, Southern University of Science and Technology, Shenzhen, China.}
}

\maketitle

\begin{abstract}
\textcolor{black}{$\emph{Objective:}$ Quantitative technique based on In-line phase-contrast computed tomography with single scanning attracts more attention in application due to the flexibility of the implementation. However, the quantitative results usually suffer from artifacts and noise, since the phase retrieval and reconstruction are independent ("two-steps") without feedback from the original data. Our goal is to develop a method for material quantitative imaging based on a priori information specifically for the single-scanning data. $\emph{Method:}$ An iterative method that directly reconstructs the refractive index decrement $\delta$ and imaginary $\beta$ of the object from observed data ("one-step") within single object-to-detector distance (ODD) scanning. Simultaneously, high-quality quantitative reconstruction results are obtained by using a linear approximation that achieves material decomposition in the iterative process. $\emph{Results:}$ By comparing the equivalent atomic number of the material decomposition results in experiments, the accuracy of the proposed method is greater than 97.2$\%$. $\emph{Conclusion:}$ The quantitative reconstruction and decomposition results are effectively improved, and there are feedback and corrections during the iteration, which effectively reduce the impact of noise and errors. $\emph{Significance:}$ This algorithm has the potential for quantitative imaging research, especially for imaging live samples and human breast preclinical studies.}

\end{abstract}


\section{Introduction}
\label{sec:introduction}
\IEEEPARstart{X}{-ray} computed tomography $($CT$)$ technology is widely used in medicine and industry. Conventional monochromatic X-ray CT imaging is based on the difference in radiation absorption between substances or tissues.
Since this imaging modality relies on a single X-ray spectrum, it is not sufficient for quantitative  imaging  of materials, especially low-Z  substances  (low-atomic-number  material,  such  as  human  soft  tissues).
Therefore, it is necessary to introduce new imaging mechanisms and technologies to meet the needs of CT material quantitative imaging.

During the last decades, dual-energy CT and phase-contrast imaging have been developing rapidly and made great progress in material quantitation \cite{a1,a2,a3,a4,a5,a6}. 
\textcolor{black}{There are some methods for material decomposition that combine spectral and phase contrast information. Mechlem \emph{et al.} confirmed that spectral grating-based phase-contrast imaging can strongly reduce the noise level of the image\cite{a7}.}
Compared with conventional CT, they have some additional requirements.
For dual-energy CT, projection data need to be acquired twice with different kilovoltage peak by using energy-integration detector or only once by using photon-counting detector (PCD) \cite{a8,a9}. For the grating-based imaging,
gratings play a crucial role in material quantitation \cite{a10,a11,a12,a13}. 
In phase-contrast imaging, it is known that the X-ray phase-shifting caused by low-Z materials at low energy is 1000 times larger than the change in its absorption value \cite{a14,a15}, which means the additional sensitivity is available.
 Many techniques can measure the phase-shifting properties of the sample, in addition to grating-based imaging \cite{a16,a17}, there are crystal interferometry \cite{a18,a19}, analyzer-based phase-contrast \cite{a20,a21} and In-line phase-contrast imaging, also known as propagation-based phase-contrast imaging (PPCI)\cite{a22,a23,a24}. Unlike other techniques, PPCI does not require other optical components so as to facilitate the implementation.

A number of propagation-based phase retrieval algorithms need to measure X-ray intensity at two or more object-to-detector distance (ODD) that may increase the radiation dose received by the object, data collection time and processing difficulty \cite{a25,a26,a27}. Meanwhile, the multi-energy propagation-based phase-contrast methods have the same disadvantages due to the multi-scans on single ODD \cite{a28,a4}. Therefore, it is necessary to develop phase retrieval algorithms based on single scanning. 
As we know, the phase retrieval problem of PPCI becomes an ill-posed inverse problem in single scanning, since it is a challenge to retrieve the phase and absorption of the sample simultaneously from single original data. Therefore, some additional assumptions or prior information are required \cite{a29,a30,a31}. 
The current single-distance phase retrieval methods use the assumption $\delta=a*\beta$, which assumes that the sample is composed of a single material. Generally, many specimens often consist of two or more materials in real application. \textcolor{black}{Therefore, there are some work for quantitative multi-material phase retrieval algorithms combined an improved assumption ${\delta _1} = {\delta _0} + \frac{{\Delta \delta }}{{\Delta \mu }}{\mu _1}$ \cite{a32}. Beltran \emph{et al.} proposed a method that retrieve the difference $({\mu _2} - {\mu _1}){T_2}$ in projected attenuation between two materials and the total projected thickness of the object in each direction need to be known\cite{a33}.
The total attenuation retrieval and correcting interface for all materials by using the improve assumption was proposed by Ulherr \emph{et al.} \cite{a34}.} 

PPCI and CT work together to obtain the $\delta$ tomography of samples \cite{a35,a36,a37,a38,a39,a40,a41}. \textcolor{black}{But it is usually implemented in two-steps: firstly, the phase shifting is retrieved from the X-ray intensity; secondly, the object is reconstructed by a conventional algorithm such as Filter Back Projection (FBP) or Algebraical Reconstruction Technique (ART) \cite{a42}. However, the reconstructed images usually suffer from artifacts and noise, since the projected phase retrieval and reconstruction are independent without feedback from the observed data.} It is widely acknowledged that one-step concepts have many advantages and are still being studied in X-ray imaging field. In multi-energy CT (incl. photon-counting CT) several groups have demonstrated that an iterative one-step reconstruction can improve quantitative results\cite{a43,a44,a45}. Some work in grating-based phase-contrast reconstruction show that it is possible to reconstruct without stepping when using a one-step algorithm\cite{a46,a47,a48,a49}.
\textcolor{black}{It is worth noting that these algorithms are related to different imaging physics and mathematical models.}  Multi-energy CT is based on the imaging principle that depends on the energy attenuation behavior of the material is related to the X-ray spectrum CT. However, this principle is challenged in the real application with low-Z compounds, which have weak absorption.
\textcolor{black}{Grating-based imaging is based on the grating self-imaging effect in optics, which also means that additional optics devices are required for imaging.} The experimental operation is complicated and the grating will reduce the utilization rate of X-ray. Hence, the authors believe that it is potential to use a one-step concept for quantitative imaging of low-Z samples without additional optical components in PPCI.

In this work, we investigate a one-step method based on single scanning, which can simultaneously reconstruct the images of the absorption factor $\beta$ and the refractive index decrement $\delta$ from original data for propagation-based phase-contrast tomography (AR-PPCT). \textcolor{black}{Meanwhile, aiming at the high-precision quantitative reconstruction of multi-material, we combine this iterative method with the improved approximation for accurate dual-material decomposition.}

\section{Theory and Methods}

\subsection{Imaging theory of PPCI}

The interaction between X-rays with matter can be described by the following complex refractive index:
\begin{equation}
n=1-\delta+i\beta.
\end{equation}

Here $\delta$ is the phase shifting factor, $\beta$ the absorption factor, and $\beta={\frac{\lambda}{{{4}{\pi}}}{\mu}}$, $\lambda$ the wavelength, $\mu$ the linear attenuation coefficient. In terms of value, $\delta$ is much larger than $\beta$. When an X-ray plane wave $A^{in}$ pass through the object, the wave function of the emergent beam reads
\begin{equation}
A(x,y)={{A}^{in}}\exp (-\frac{M(x,y)}{2}+\mathbf{i}\Phi (x,y)),    
\end{equation}
where $M(x,y)$ and $\Phi (x,y)$ are the X-ray absorption and phase shift, respectively.
\begin{equation}
M(x,y)=\frac{4\pi }{\lambda }\int_{l}{\beta (x,y,z)dz},  
\end{equation}   
\begin{equation}
\Phi (x,y)=-\frac{2\pi }{\lambda }\int_{l}{\delta (x,y,z)dz},
\end{equation}
here $l$ is the integration path. After penetrating the object, the intensity of X-ray decays to
\begin{equation}
{{I}_{out}}(x,y)={{\left| A(x,y) \right|}^{2}}={{I}^{in}}\exp (-M(x,y)),
\end{equation}
here ${{I}^{in}}={{\left| {{A}^{in}} \right|}^{2}}$ is the intensity of incoming X-rays.

In the case of paraxial approximation, according to the Fresnel diffraction theory, the plane X-ray intensity distribution at the distance $z$ from the sample can be written as \cite{a50} :
\begin{equation}
{{I}_{z}}(x,y)={{\left| {{h}_{z}}\otimes A(x,y) \right|}^{2}},
\end{equation}
where $\otimes$ represents convolution, ${{h}_{z}}$ is the Fresnel propagator:
\begin{equation}
{{h}_{z}}=\frac{\exp (\mathbf{i}kz)}{\mathbf{i}\lambda z}\exp (\mathbf{i}\frac{\pi }{\lambda z}({{x}^{2}}+{{y}^{2}})),
\end{equation}
here $k=\frac{2\pi }{\lambda }$ is the wave number.

The mathematical problem of PPCT is  that reconstruct the  $\bm{\delta}$ and $\bm{\beta}$  of the measured sample from a serial of  intensity $I_z$ with different paths $l$.

\subsection{The linear relationship approximation}
By single scanning in propagation-based phase contrast imaging, the phase retrieval problem becomes an ill-posed inverse problem, since there is only one set of intensity data and two unknown variables ($\delta$ and $\beta$). It is necessary to establish a certain hypothetical relationship between $\delta$ and $\beta$.
 One of the most widely used assumptions in phase retrieval within single ODD scanning is a multiple relationship: ${\delta }=a*{{\beta }}$,  which means that the object consists of only a single material. However, the great majority of objects often consist of more than one material in application. How to achieve high-precision value of $\delta$ and $\beta$ that become very significant for research in PPCI with single ODD.

Firstly, we suppose whether we can define a higher-order approximation to the $\delta$ and $\beta$ of the multi-base materials. The
approximation is expressed as follows:
\begin{equation}
\label{equ:decom00}
    \delta  = {a_N}{\beta ^N} + {a_{N - 1}}{\beta ^{N - 1}} + {a_{N - 2}}{\beta ^{N - 2}} +  \ldots  + {a_1}\beta  + {a_0},
\end{equation} 
the above formula is an expression of an equation of degree $\bm{N}$ in one variable $\bm{\beta}$.

\textcolor{black}{Dual-material decomposition, which allows quantitative material images and beamhardening artifact reduction, has many applications in medicine and biology. In this case, the approximation is simple.} \textcolor{black}{It can be expressed as:
\begin{equation}
\label{equ:decom0}
{{\delta }_{i}}=a{{\beta }_{i}}+b,\text{   }( a>0, i=1,2),
\end{equation}
here ${{\beta }_{i}}$ and ${{\delta }_{i}}$ represent the absorption factor and the phase-shifting factor of the $i$th base material. This equation means that the case of $N=1$ in Eq.~(\ref{equ:decom00})}. When the type of sample material is known, $a$ and $b$ can be fitted by the least square method:
\begin{equation}
min \sum\limits_{i = 1}^t {{{\left[ {{\delta _i} - \left( {a{\beta _i} + b} \right)} \right]}^2}},
\end{equation}

\begin{figure}[t]
	\centerline{\includegraphics[width=\columnwidth]{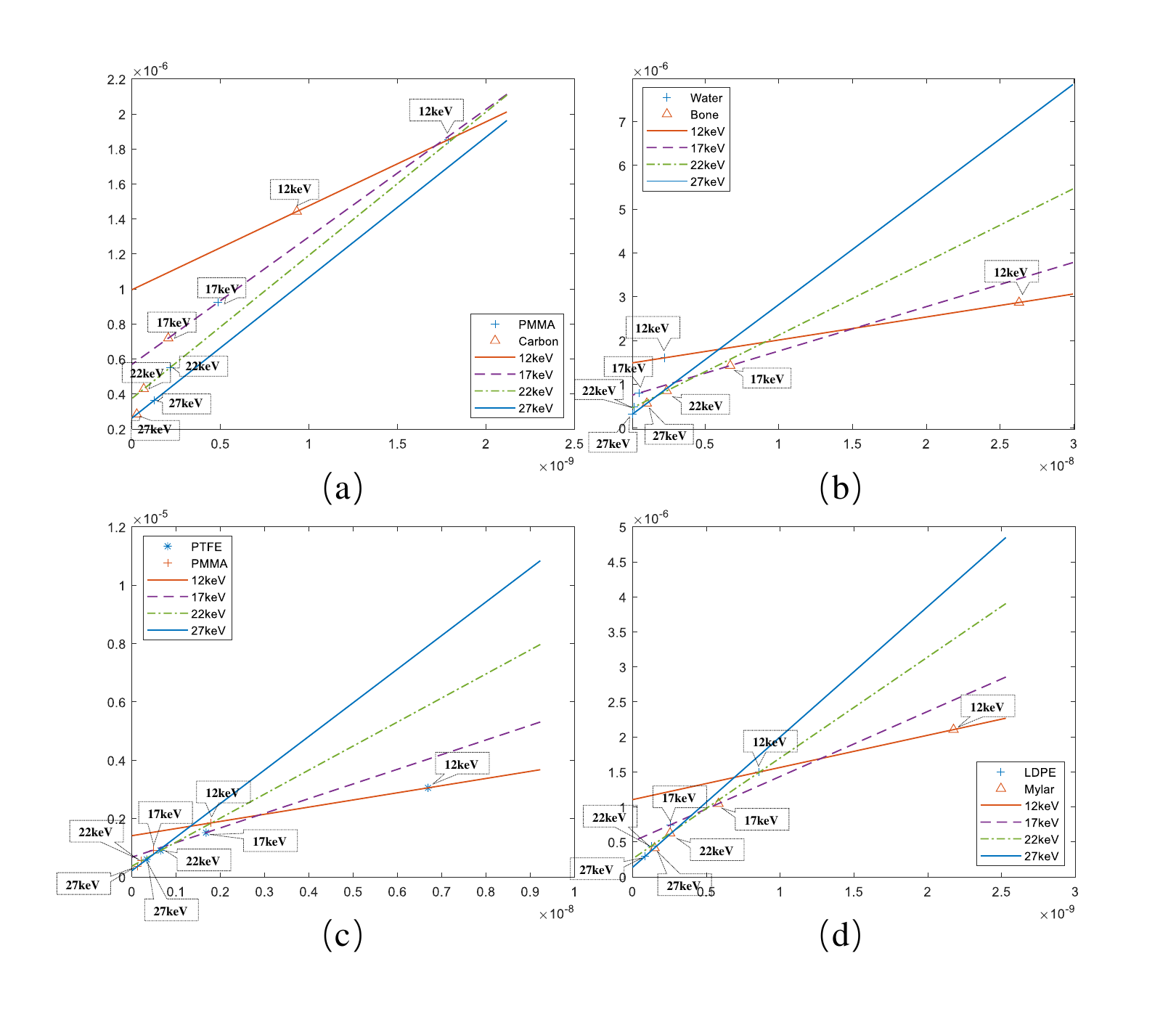}}
	\caption{Geometric illustration of the linear relationship between different base material pairs in the phase-contrast energy range (12keV-27keV). }
	\label{Fig.1}
\end{figure}

\textcolor{black}{Figure.\ref{Fig.1} gives a geometric illustration of the linear approximation $\delta=a\beta+b$ about different base material pairs in the normal phase-contrast energy range.
The selected materials are well-known, which are namely Polytetrafluoroethylene (PTFE), Polymethyl Methacrylate (PMMA), Low Density Polythylene (LDPE), Carbon, Mylar, Bone and Water. We choose material pairs with different atomic numbers. The horizontal and vertical coordinates represent $\beta$ and $\delta$ of the material, respectively. As shown in Fig.\ref{Fig.1}, it is obvious that the two base materials can be perfectly fitted with this approximation in the current energy range.
After analysis and verification, the linear relationship is valid for different substances, which makes possible for material decomposition for PPCI in single energy and scanning.} 

\textcolor{black}{\textcolor{black}{According to some conversion formulas for material decomposition in phase-contrast imaging\cite{a4,a5,a10}, the $\bm{\beta}$ and $\bm{\delta}$ of the sample can be expressed as a linear combination of $\beta$ and $\delta$ of the two basic materials, respectively . } 
The decomposition of the sample can be carried out:
\begin{equation}
\label{equ:decom1}
\left\{ \begin{array}{*{35}{l}}
\beta(x,y)=f(x,y){{\beta }_{1}}+g(x,y){{\beta }_{2}}  \\
\delta(x,y)=f(x,y){{\delta }_{1}}+g(x,y){{\delta }_{2}}  \\
\end{array} \right.,
\end{equation}
$f(x,y)$ and $g(x,y)$ are the distribution function of the two substrates, ${{\beta }_{i}}$ and ${{\delta }_{i}}(i=1,2)$ are the absorption factor and the phase-shifting factor of the sample-based material, respectively.}

\textcolor{black}{Two different material combinations can achieve the corresponding decomposition with this linear relationship. Therefore, after preselecting dual-material, we pay attention to material decomposition and quantitation for multi-material samples.}

\subsection{Reconstruction algorithm}
According to the Born approximation, the X-ray intensity ${{I}_{z}}$ in the image plane ($z$ represents the distance from the sample to the image plane) satisfies the following mathematical relationship \cite{a51}:
\begin{equation}
\label{equ:de2}
{{\mathcal{F}\left[ \frac{\frac{{{I}_{z}}}{{{I}_{0}}}-1}{2} \right]}}=\\
\cos (z{{\rho }^{2}}){{\operatorname{Re}{{\psi }_{0}}}}+\sin (z{{\rho }^{2}}){{\operatorname{Im}{{\psi }_{0}}}},
\end{equation}
here ${\mathop{\rm Re}\nolimits} {\psi _0} = \mathcal{F}[ - k\int {\beta dl} ]$, ${\mathop{\rm Im}\nolimits} {\psi _0} = \mathcal{F}[ - k\int {\delta dl} ]$. ${{\rho }^{2}}=\pi \lambda ({{\zeta }^{2}}+{{\eta }^{2}})$, \textcolor{black}{$(\zeta ,\eta )$ are the frequency domain coordinates of $(x,y)$.
$\mathcal{F}$ is the Fourier transform}. 

\textcolor{black}{Let $\bm{\beta} {\rm{ = (}}{\beta _1}{\rm{,}}{\beta _2}, \ldots {\beta _J}{{\rm{)}}^\tau }$ and $\bm{\delta} {\rm{ = (}}{\delta _1}{\rm{,}}{\delta _2}, \ldots {\delta _J}{{\rm{)}}^\tau }$ denote the discretized images of $\beta(x,y)$ and $\delta(x,y)$, where $\beta_j$ and $\delta_j$ are the sampled values of $\beta(x,y)$ and $\delta(x,y)$ at the $j$th pixel, $J$ the total pixel number, and $\tau$ the vector transpose operation. ${R^\varphi } = {(r_{uj}^\varphi )_{U \times J}}$ is the projection matrix at angle $\varphi$, where $(r_{uj}^\varphi )$ represents the contribution of $\delta_j$ and $\beta_j$ to the projection along the $u$-th x-ray path at projection angle $\varphi$. \textcolor{black}{$I_z^\varphi$ is U dimensional column vector}, $U$ is the number of detector cell.
Firstly, we obtain the residual X-ray intensity by subtracting the measured intensity and the simulated intensity. the intensity residual of the $m$-th iteration at projection angle $\varphi$ is:
\begin{equation}
\operatorname{Re}_{z}^{\varphi(m)}={{I}_{z}^{\varphi}}-{{\left| {{h}_{z}}\otimes\left[ {{A}^{in}}\exp (-\frac{{{M}^{\varphi(m)}}}{2}+\mathbf{i}{{\Phi }^{\varphi(m)}}) \right] \right|}^{2}},
\end{equation}
here ${{M}^{\varphi(m)}}=2k\int{{{\bm{\beta}}^{\varphi(m)}}}dl$, and ${{\Phi}^{\varphi(m)}}=-k\int{{{\bm{\delta}}^{\varphi(m)}}}dl$.} 

\textcolor{black}{Utilizing the linear relation Eq.~(\ref{equ:decom0}), we can attain ${\mathop{\rm Im}\nolimits} {\psi _0} = a{\mathop{\rm Re}\nolimits} {\psi _0} + \mathcal{F}[ - k\int b dl]$. And the absorption residual of the \textcolor{black}{$m$}-th iteration at projection angle $\varphi$ is:
\begin{equation}
    {\mathop{\rm Re}\nolimits} {\psi ^\varphi } - {\mathop{\rm Re}\nolimits} {\psi ^{\varphi (m)}} = \frac{{\mathcal{F}({\mathop{\rm Re}\nolimits} _z^{\varphi (m)}/2)}}{{\cos (z{\rho ^2}) + \sin (z{\rho ^2})*a}},
\end{equation}}
\textcolor{black}{here $\mathcal{F}$ is 1D Fourier transform, and ${\mathop{\rm Re}\nolimits}={{\mathcal{F}}^{ - 1}}({\mathop{\rm Re}\nolimits} {\psi ^\varphi } - {{\mathop{\rm Re}\nolimits} ^{\varphi (m)}})$ is U dimensional column vector}, \textcolor{black}{${{\mathcal{F}}^{-1}}$ is 1D Inverse Fourier transform. Then combining the Simultaneous Algebraical Reconstruction Technique (SART)\cite{a52}, we can directly reconstruct  $\bm{\delta}$ and  $\bm{\beta }$ of the sample for \textcolor{black}{m+1} iterations, the scheme is as follows:}
\textcolor{black}{
\begin{equation}
\label{equ:de0}
\left\{ \begin{array}{*{20}{l}}
{\bm{\beta} _j^{m + 1} = \bm{\beta} _j^m + \frac{\gamma }{{R_{ + ,j}^\varphi }}\sum\limits_{u = 1}^U {\frac{{r_{u,j}^\varphi }}{{R_{u, + }^\varphi }}[( - {k^{ - 1}}){{{{\mathop{\rm Re}\nolimits}}_u}}]} }\\
{\bm{\delta} _j^{m + 1} = \bm{\beta} _j^{m + 1}*a + b}\\
\end{array} \right.,
\end{equation}}
\textcolor{black}{where $R_{u, + }^\varphi  = \sum\nolimits_{j = 1}^J {r_{u,j}^\varphi }$ with $u = 1,2, \ldots U$, and $R_{ + ,j}^\varphi  = \sum\nolimits_{u = 1}^U {r_{u,j}^\varphi }$ with $j = 1,2, \ldots J$. $\gamma$ the relaxation factor, which can be obtained by simulating the real reconstruction environment experimentally, and then selecting the relaxation factor with the best reconstruction effect and applying it to the real experiment. The above formulas represent the reconstruction of $\bm{\delta}$ and $\bm{\beta}$ based on the projection angle $\varphi$, and we need to reconstruct the tomography from multiple different projection angles.}

\textcolor{black}{The $shrink$ function is a mask to remove the non-zero value of the air in the reconstructed image:  
\begin{equation}
\label{equ:de1}
shrink(y,x)=\left\{ \begin{matrix}
y=0  \\
y=y  \\
\end{matrix} \right.\text{                 }\begin{matrix}
(b-x <y<b+x )  \\
else  \\
\end{matrix},
\end{equation}
here $y$ is the reconstructed image and $x$ is the parameter. When $y$ is the $\bm{\delta}$ image, the value of $x$ is less than $100$ times that of the constant term $b$ in Eq.~(\ref{equ:decom0}).}

\textcolor{black}{Finally, we summarize the implementation steps of the algorithm:}

\begin{algorithm}[htb] 
	\caption{ \textcolor{black}{The AR-PPCT algorithm.}} 
	\begin{algorithmic}[1] 
		
		\STATE Initialization ${{\bm{\beta}}^{0}}=0,\text{ }{{\bm{\delta}}^{0}}=0,\text{ $m$=0}$; 
		
		\STATE Use Eq. $(6)$-$(7)$  to calculate the intensity $I_{z}^{\varphi(m)}$ , and gain the residual intensity  $\operatorname{Re}_{z}^{\varphi(m)}$; 
		
		\STATE Iteratively update ${{\bm{\beta}}^{m+1}}$ and ${{\bm{\delta}}^{m+1}}$ according to Eq.~(\ref{equ:de0}); 
		\STATE Use Eq.~(\ref{equ:de1}) to remove the non-zero value of air in the image, and then material decomposition by Eq.~(\ref{equ:decom1});
		
		\STATE Set $m=m+1$ and turn to step 2 until the stop condition is met; 
		
		\STATE Return ${{\bm{\beta}}^{m}}$ and ${{\bm{\delta}}^{m}}$; 
		
	\end{algorithmic}
\end{algorithm}

\section{Experiments}

\textcolor{black}{In this section, the proposed algorithm is evaluated by numerical simulations and real experiments. As a comparison, we have also tested the Born approximation method with single ODD\cite{a53}, Linear method\cite{a34} and Material Decomposition using Spectral propagation-based phase imaging (MD-SPBI) method \cite{a28} in experiments. In this paper, three comparison algorithms use SART algorithms for reconstruction after phase and absorption/decomposed materials projection is retrieved.} \textcolor{black}{ We devise stopping criteria in mathematics is $\frac{{\left\| {{\mathop{\rm Re}\nolimits} \psi  - {\mathop{\rm Re}\nolimits} {\psi ^m}} \right\|}}{{\left\| {{\mathop{\rm Re}\nolimits} \psi } \right\|}} \le \varepsilon$, here $\varepsilon  \to 0$.}

\textcolor{black}{According to the original reference, MD-SPBI, which uses a set of basis functions defined by the specific materials, can obtain the base material decomposition projection image. The Born approximation with a single ODD and Linear method are phase retrieval algorithms without material decomposition.}The $\delta=a\beta$ is used for Born approximation with a single ODD, \textcolor{black}{and the $\delta=a\beta+b$ is used for the Linear method}. The parameter $a$ in  ${\delta }=a{{\beta }}$ is fitted by the least square method. The phase projection can be obtained by this multiple relationship after calculating the absorption projection in Born approximation single-distance method. Meanwhile, the $\bm{\delta}$ image is obtained using this linear relationship again after removing the artifacts of $\bm{\beta}$ image in the linear method. \textcolor{black}{After getting $\bm{\beta}$ and $\bm{\delta}$ images, we use Eq.~(\ref{equ:decom1}) to get the decomposition \textcolor{black}{results of the methods.}} \textcolor{black}{Furthermore, since MD-SPBI is a multi-energy method, the second energy should be chosen in experiments. We refer to the energy value range in the thesis of MD-SPBI method for the following two reasons: 1) An energy range determined according to imaging requirements and actual conditions; 2) The specific value takes into account the energy value of the comparison methods.}

\subsection{Numerical Simulation}
\begin{figure}[!t]
	\centerline{\includegraphics[width=\columnwidth]{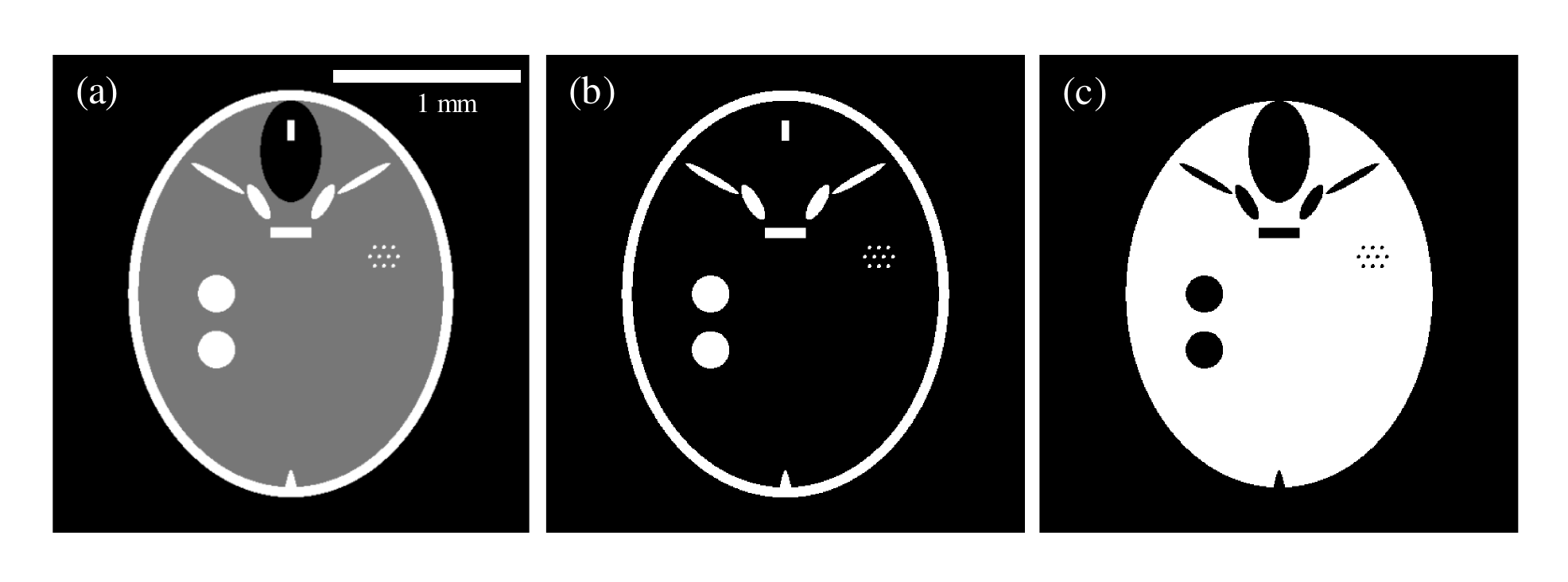}}
	\caption{Phantom utilized in the numerical experiments. (a)Numerical phantom; (b)The image of bone component; (c)The image of water component. }
	\label{Fig.2}
\end{figure}
\begin{figure}[!t]
	\centerline{\includegraphics[width=\columnwidth]{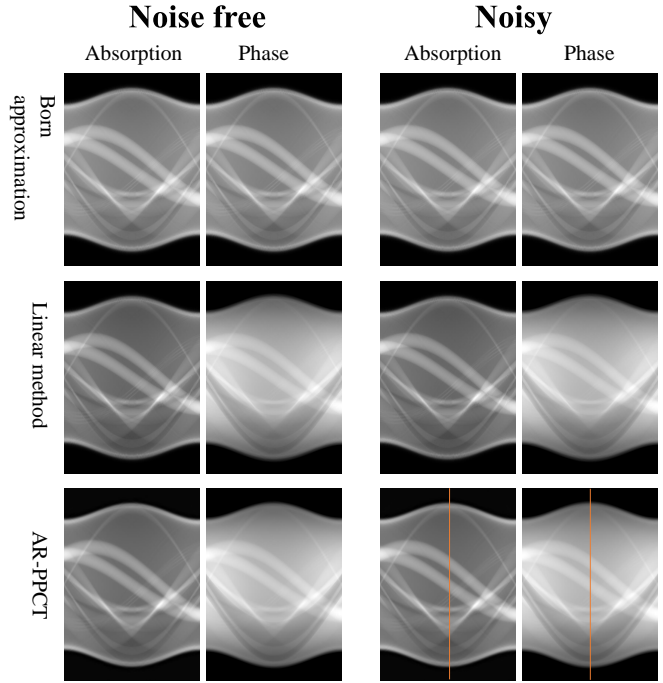}}
	\caption{The retrieved projections of absorption and phase in noise-free case and noisy case. }
	\label{Fig.3}
\end{figure}
\begin{figure}[!t]
	\centerline{\includegraphics[width=\columnwidth]{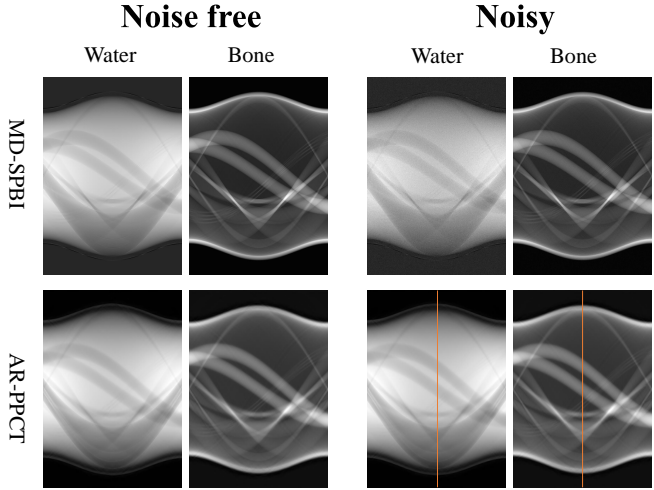}}
	\caption{The retrieved projections of water and bone in noise-free case and noisy case. }
	\label{Fig.4}
\end{figure}

\begin{figure}[!t]
	\centerline{\includegraphics[width=\columnwidth]{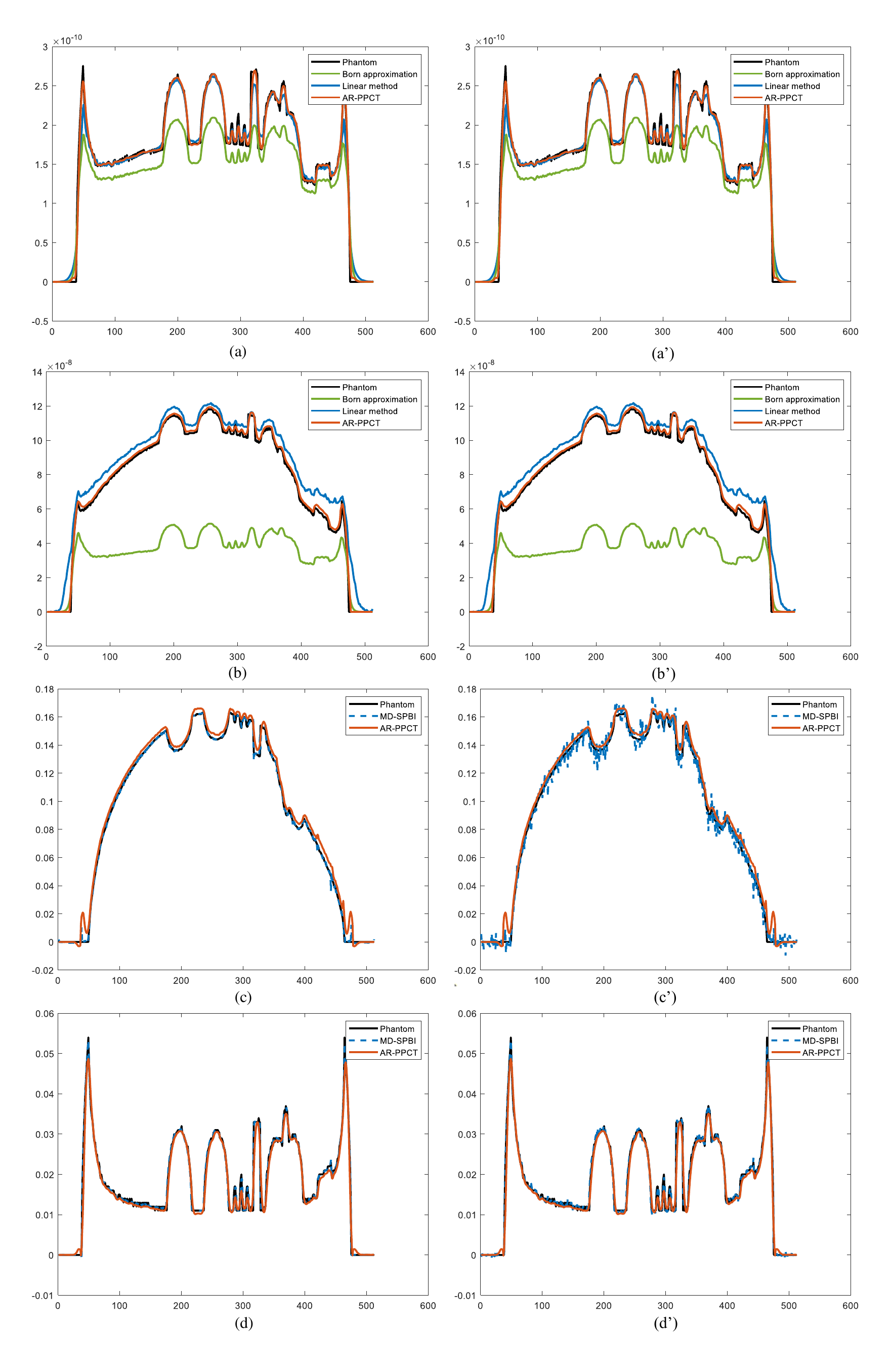}}
	\caption{Profiles of projections in noise-free case and noisy case. (a) and (b) are the profile lines of absorption and phase projections in noise-free case. (c) and (d) are the profile lines of water-based and bone-based material in noise-free case. (a'), (b'), (c') and (d') are the profile lines of absorption, phase, basic materials in noisy case, respectively. }
	\label{Fig.5}
\end{figure}

\begin{figure}[!t]
	\centerline{\includegraphics[width=\columnwidth]{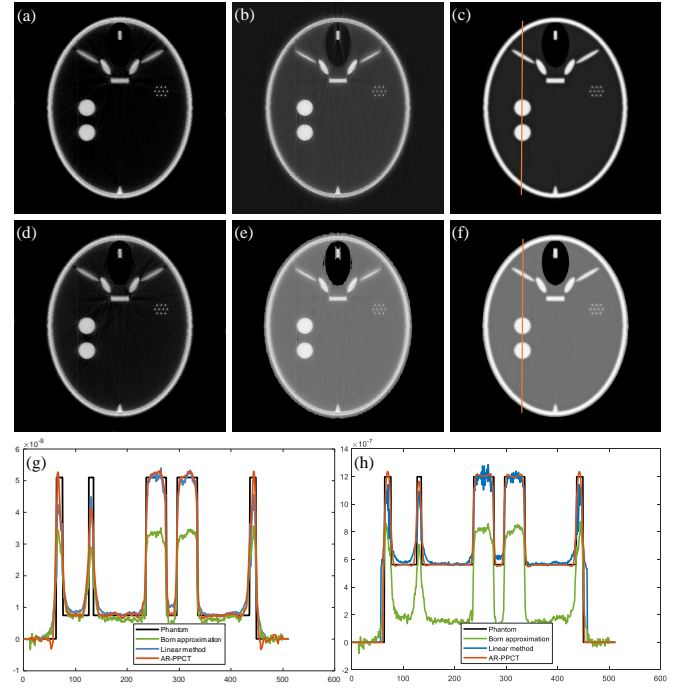}}
	\caption{The tomographies in noise-free case. (a)The $\bm{\beta}$ image of Born approximation with single ODD method; (b)The $\bm{\beta}$ image of Linear method; (c)The $\bm{\beta}$ image of AR-PPCT; (g)The profile lines of $\bm{\beta}$ images; (d), (e) and (f) are the $\bm{\delta}$ images of Born approximation with single ODD method, linear method and AR-PPCT respectively; (h)The profile lines of $\bm{\delta}$ image. }
	\label{Fig.6}
\end{figure}
\textcolor{black}{As shown in Fig.\ref{Fig.2}, a  FORBILD head phantom was utilized in the numerical experiment \cite{a54}. The experimental parameters are displayed in Table \ref{table.10}. In the simulation, a parallel-beam setting was used for acquiring 360 projections equally spaced in 180 degrees. The size of sample was 2.45 mm*1.95 mm. We simulated both noise free data and Poisson noise data corresponding to emission flux of $\text{1}{{\text{0}}^{\text{6}}}$ photons per measurement. The water and bone were selected as the basic materials for quantitative imaging. The values of $\delta$ and $\beta$ were from the X-ray database provided by \underline{http://henke.lbl.gov/optical$\_$constants/getdb2.html}.  The size of reconstructed image was 512*512, and the maximum iteration was set as \textcolor{black}{200} which was the stop condition of AR-PPCT. We chose the second energy as 25 keV in MD-SPBI.}
\begin{table}[]
	\caption{{Experimental parameters in numerical simulation}.}
	\centering
	\scalebox{1.0}{
		\begin{tabular}{cc}
			 \toprule 
			
			\multirow{1}{*}{Energy}&18 keV\\
		
		    \multirow{1}{*}{ODD}&20 cm\\
			\multirow{1}{*}{Pixel Array Detector}&512*1\\
			
			\multirow{1}{*}{Detector Unit Size} &5 um \\

			\bottomrule
			\label{table.10}
	\end{tabular}}
\end{table}

\begin{figure}[!t]
	\centerline{\includegraphics[width=\columnwidth]{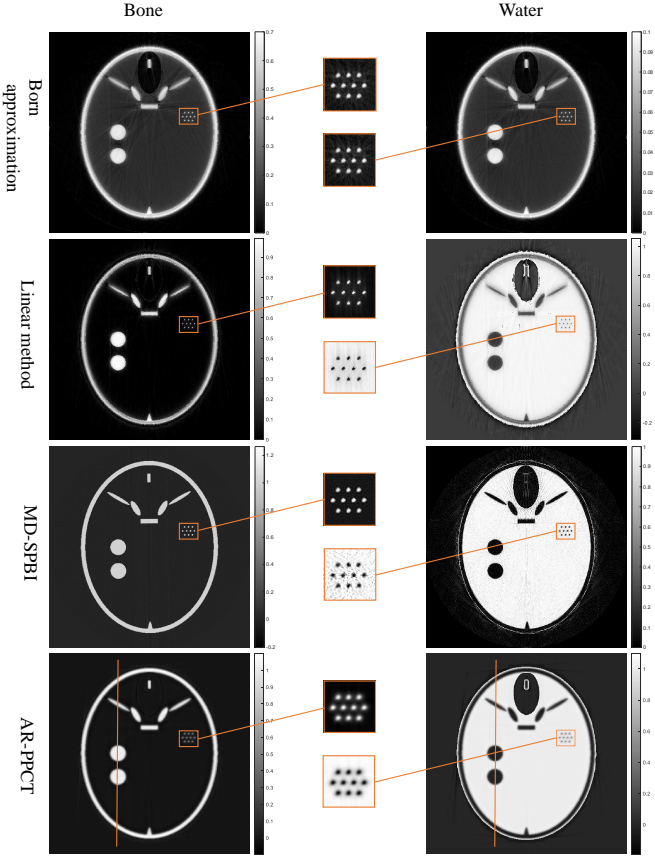}}
	\caption{The decomposed results of the model in noise-free case. }
	\label{Fig.7}
\end{figure}

\begin{figure}[!t]
	\centerline{\includegraphics[width=\columnwidth]{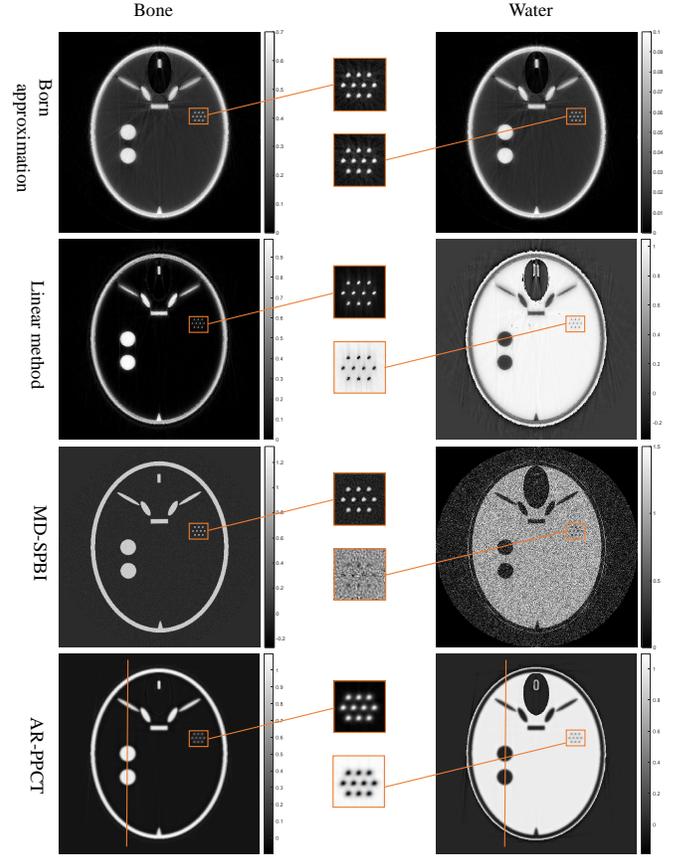}}
	\caption{The decomposed results of the model in noisy case. }
	\label{Fig.8}
\end{figure}

\textcolor{black}{Fig.\ref{Fig.3}-\ref{Fig.5} show the retrieved projections and the profiles of the projections in noise-free and noisy cases, respectively. It is noted that absorption and phase, water and bone projection maps are obtained by Radon transformation in AR-PPCT. The tomographies of $\bm{\beta}$ and $\bm{\delta}$ in noise-free are shown in Fig.\ref{Fig.6}.}
\textcolor{black}{The decomposed results of noise-free and noisy cases are shown in Fig.\ref{Fig.7} and Fig. \ref{Fig.8}, respectively. For each case, the results of Born approximation with a single ODD, Linear method, MD-SPBI method and our method are performed. The Peak Signal to Noise Ratio (PSNR) and Structural Similarity (SSIM) of water-based and bone-based materials for these methods in two cases are shown in Table \ref{table.2} and \ref{table.3}.} 

\textcolor{black}{From the results in Fig.\ref{Fig.3}-Fig.\ref{Fig.5}, it can be seen that the absorption and phase projection results of Born approximation with single ODD are far from the ideal results due to the unsatisfactory assumption  ${\delta }={a}{{\beta }}$. The MD-SPBI method can better separate water and bone materials because of using two sets of different X-ray intensity data, which makes the phase retrieval problem in single distance of PPCI a well-posed inverse problem. Moreover, by comparing the decomposition results of water materials, it is apparent that the anti-noise ability of the MD-SPBI method is weaker than that of AR-PPCT. From Fig. \ref{Fig.5}(b,c), we also find that the profiles of phase and water projections in AR-PPCT and phase projections in Linear method are slightly higher than that of phantom, since the non-zeros value of the air in $\bm{\delta}$ image isn't removed completely.}

\begin{figure}[!t]
	\centerline{\includegraphics[width=\columnwidth]{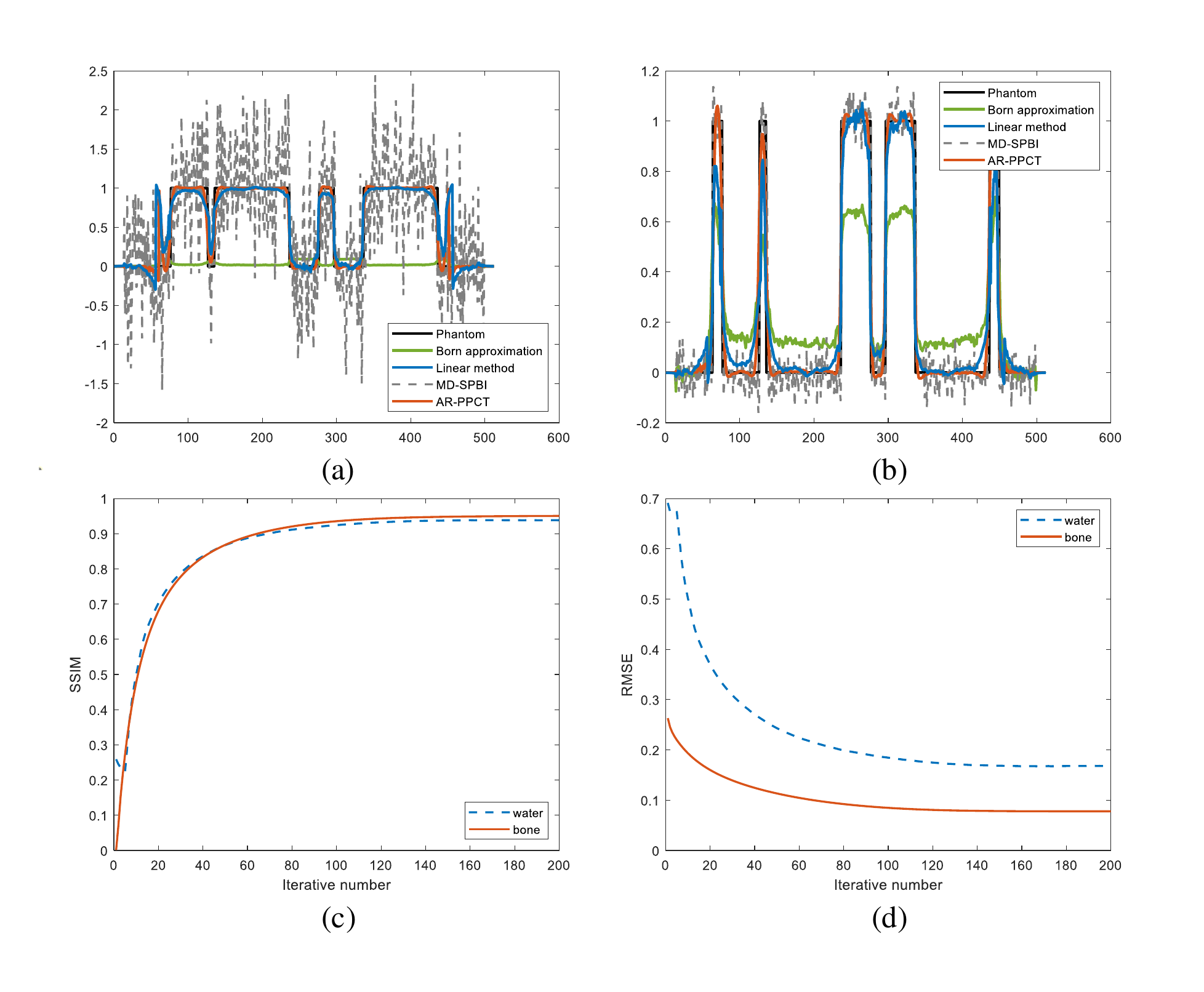}}
	\caption{The profile line of decomposed results in noisy case. (a)The profile lines of water-based material results; (b)The profile lines of bone-based material results; (c) and (d) are the SSIM and RMSE of water-based and bone-based material images in AR-PPCT. }
	\label{Fig.9}
\end{figure}

\textcolor{black}{In the noise-free case, the results are depicted in Fig.\ref{Fig.6}. Compared with the value of phantom, one obvious finding is that the $\bm{\beta}$ images of AR-PPCT and Linear method are superior to that of Born approximation method with single ODD. The $\bm{\delta}$ result of AR-PPCT is better than that of Born approximation with single ODD because of the improved assumption ${\delta }=a{{\beta }+b}$ and there are some artifacts in Linear method. Material decomposition can be performed after obtaining $\bm{\delta}$ and $\bm{\beta}$  by Eq.~(\ref{equ:decom1}). The decomposed results are shown in Fig.\ref{Fig.7}, it is clear that Linear method, MD-SPBI and AR-PPCT can decompose water-based and bone-based materials including zoom area. In the noisy case, looking at Fig.\ref{Fig.8}, AR-PPCT, Linear method and Born approximation with single ODD have certain anti-noise performance. In contrast, MD-SPBI is more sensitive to noise than other methods. Moreover, In Fig. \ref{Fig.9}(a-b), the profiles of decomposed results from orange lines in noisy case show that AR-PPCT, compared with the other methods, has a superior image quality in material quantitation.}

\textcolor{black}{The results of PSNR and SSIM, as shown in Table \ref{table.2} and \ref{table.3}, indicate that the AR-PPCT has more advantages in material decomposition.} \textcolor{black}{
The curves of the Root Mean Squard Error (RMSE) and SSIM in Fig. \ref{Fig.9}(c-d) indicate the error tends to stabilize as the number of iterations increases. Therefore, the convergence of AR-PPCT is also proved in numerical.}. 

\begin{table*}[]
	\caption{{PSNR and SSIM comparison of reconstructed water based image results}.}
	\centering
	\scalebox{0.8}{
		\begin{tabular}{ccccccccc}
			\toprule 
			&\multicolumn{3}{c}{\textbf{{Noise free}}}&\multicolumn{3}{c}{\textbf{{Noisy}}}\\
			\cmidrule(r){2-5}\cmidrule(r){6-9}
			&Born approximation&Linear method&MD-SPBI&AR-PPCT&Born approximation&Linear method&MD-SPBI&AR-PPCT\\
			\cmidrule(r){1-5}\cmidrule(r){6-9}
			\multirow{1}{*}{PSNR}&4.55&\textcolor{black}{14.01}&14.11&$\bm{15.50}$&4.55&\textcolor{black}{14.01}&6.10&$\bm{15.50}$ \\
			\cmidrule(r){1-5}\cmidrule(r){6-9}
			SSIM&0.002&\textcolor{black}{0.91}&0.92&$\bm{0.94}$&0.002&\textcolor{black}{0.91}&0.65&$\bm{0.94}$ \\
			\bottomrule
			\label{table.2}
	\end{tabular}}
\end{table*}

\begin{table*}[]
	\caption{{PSNR and SSIM comparison of reconstructed bone based image results}.}
	\centering
	\scalebox{0.8}{
		\begin{tabular}{ccccccccc}
			\toprule
			&\multicolumn{3}{c}{\textbf{{Noise free}}}&\multicolumn{3}{c}{\textbf{{Noisy}}}\\
			\cmidrule(r){2-5}\cmidrule(r){6-9}
			&Born approximation&Linear method&MD-SPBI&AR-PPCT&Born approximation&Linear method&MD-SPBI&AR-PPCT\\
			\cmidrule(r){1-5}\cmidrule(r){6-9}
			\multirow{1}{*}{PSNR}&16.11&\textcolor{black}{18.64}&23.48&$\bm{22.17}$&16.11&\textcolor{black}{18.64}&22.01&$\bm{22.17}$ \\
			\cmidrule(r){1-5}\cmidrule(r){6-9}
			SSIM&0.70&\textcolor{black}{0.86}&0.97&$\bm{0.95}$&0.70&\textcolor{black}{0.86}&0.95&$\bm{0.95}$ \\
			\bottomrule
			\label{table.3}
	\end{tabular}}
\end{table*}

\subsection{Real experiment 1}

\begin{figure}[!t]
	\centerline{\includegraphics[width=\columnwidth]{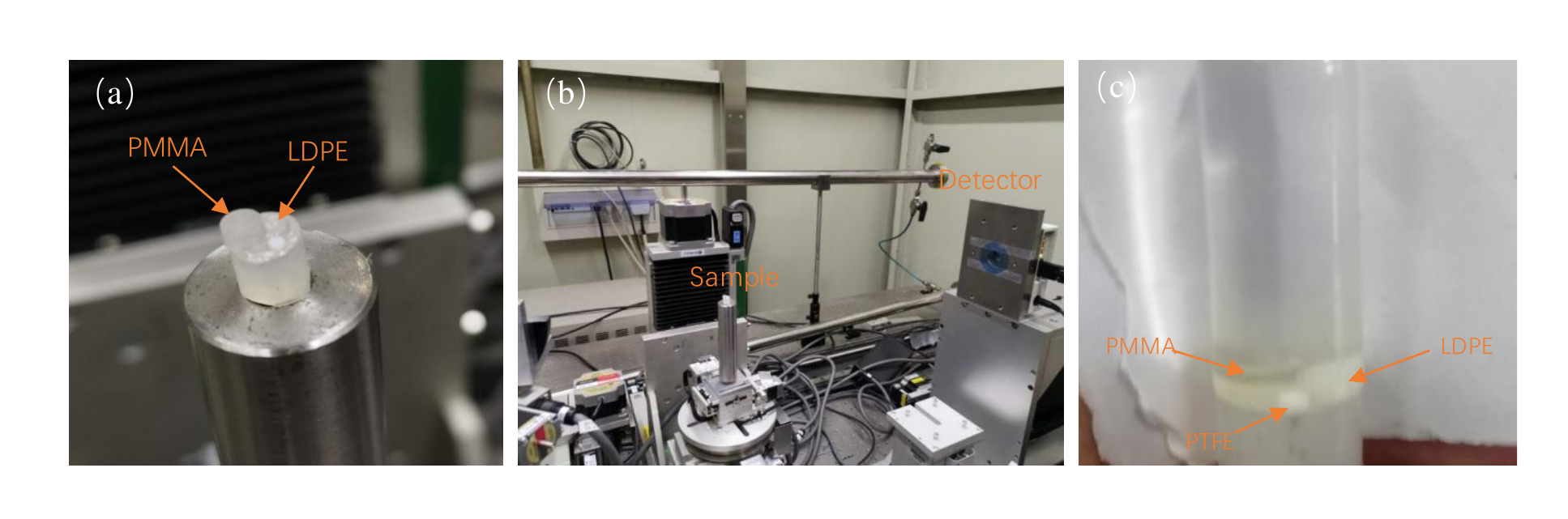}}
	\caption{The physical and facility photos. (a) The physical object of real experiment 1 in BSRF; (b) The facility of BSRF; (c)The physical object of real experiment 2 in SSRF. }
	\label{Fig.10}
\end{figure}

\textcolor{black}{The tests were carried out at the beamline 4W1A at the Beijing Synchrotron Radiation Facility (BSRF). As shown in Fig. \ref{Fig.10}(a-b), the experimental samples include PMMA and LDPE. The experimental parameters are displayed in Table \ref{table.20}.
The diameters of PMMA and LDPE were 4 mm and 2 mm. The data of 720 angles were collected at equal intervals within 180 degrees. The reconstructed image size was 512*512. We chose the second energy as 10 keV in MD-SPBI.}

\begin{table}[]
	\caption{{The experimental parameters in the Beijing Synchrotron Radiation Facility} (4W1A).}
	\centering
	\scalebox{0.8}{
		\begin{tabular}{cc}
			\toprule 
			
			\multirow{1}{*}{Energy}&15keV\\
			
			\multirow{1}{*}{ODD}&43 cm\\
			
			\multirow{1}{*}{Pixel Array Detector}&2048*2048\\
			
			\multirow{1}{*}{Detector Unit Size} &6.5 um \\
			
			\multirow{1}{*}{Single Exposure Time} &30 ms\\
			
			\multirow{1}{*}{Samples $\&$ Size} &PMMA: 4 mm; LDPE: 2 mm \\

			\bottomrule
			\label{table.20}
	\end{tabular}}
\end{table}

\begin{figure}[!t]
	\centerline{\includegraphics[width=\columnwidth]{102.pdf}}
	\caption{Projections results. (a), (b) and (c) are the retrieved absorption maps of Born approximation with single ODD method, linear method and AR-PPCT, respectively; (d), (e) and (f) are the retrieved phase maps of Born approximation with single ODD method, linear method and AR-PPCT, respectively; (g) and (h) are the profile lines of absorption and phase maps, respectively.}
	\label{Fig.11}
\end{figure}
\begin{figure}[!t]
	\centerline{\includegraphics[width=\columnwidth]{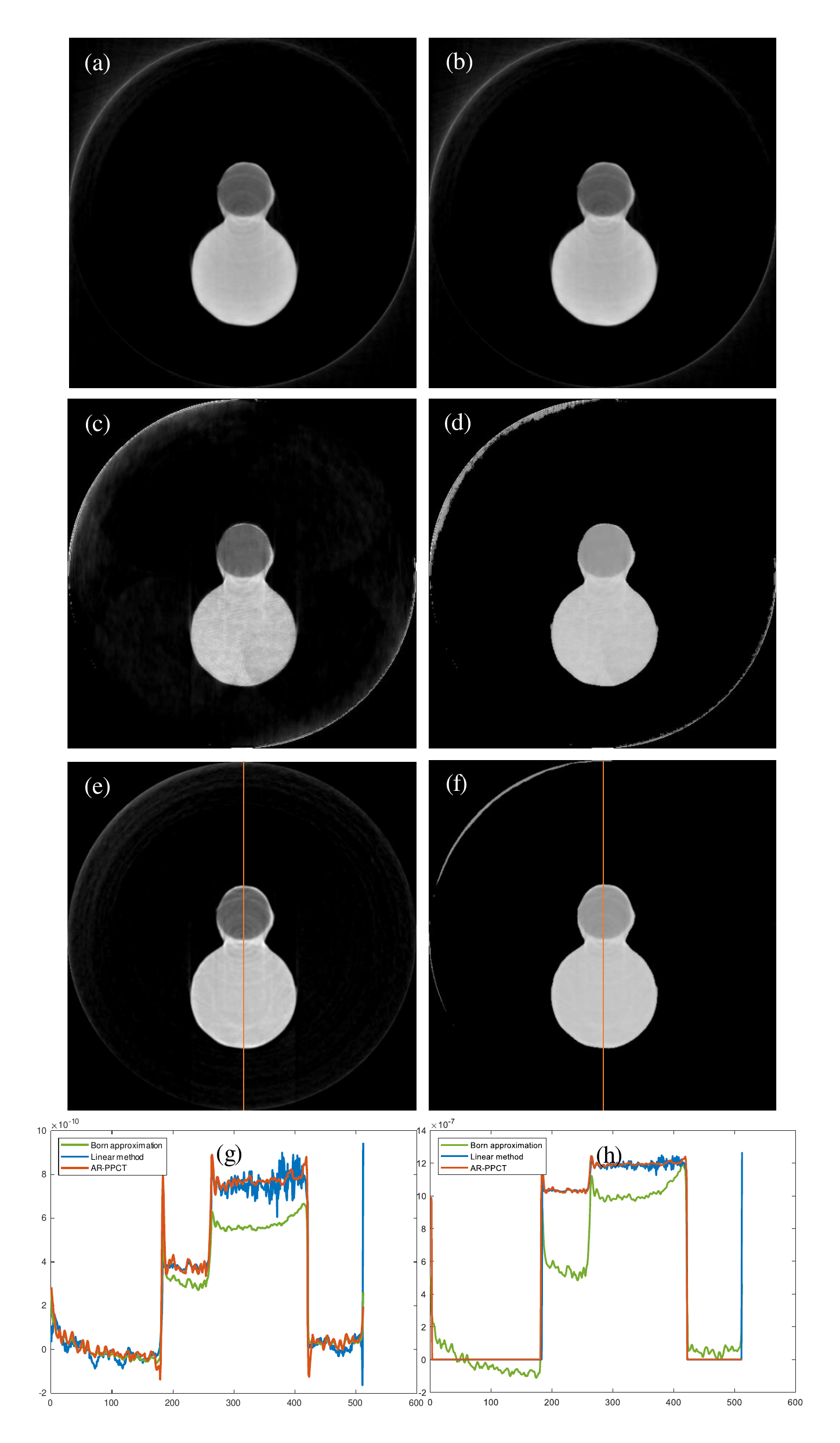}}
	\caption{Tomography results. (a)The $\bm{\beta}$ image of Born approximation with single ODD method; (c)The $\bm{\beta}$ image of Linear method; (e)The $\bm{\beta}$ image of AR-PPCT; (g)The profiles of $\bm{\beta}$ images; (b), (d) and (f) are the $\bm{\delta}$ image of Born approximation with single ODD method, linear method and AR-PPCT, respectively; (h)The profile lines of $\bm{\delta}$ images. }
	\label{Fig.12}
\end{figure}

\begin{figure}[!t]
	\centerline{\includegraphics[width=\columnwidth]{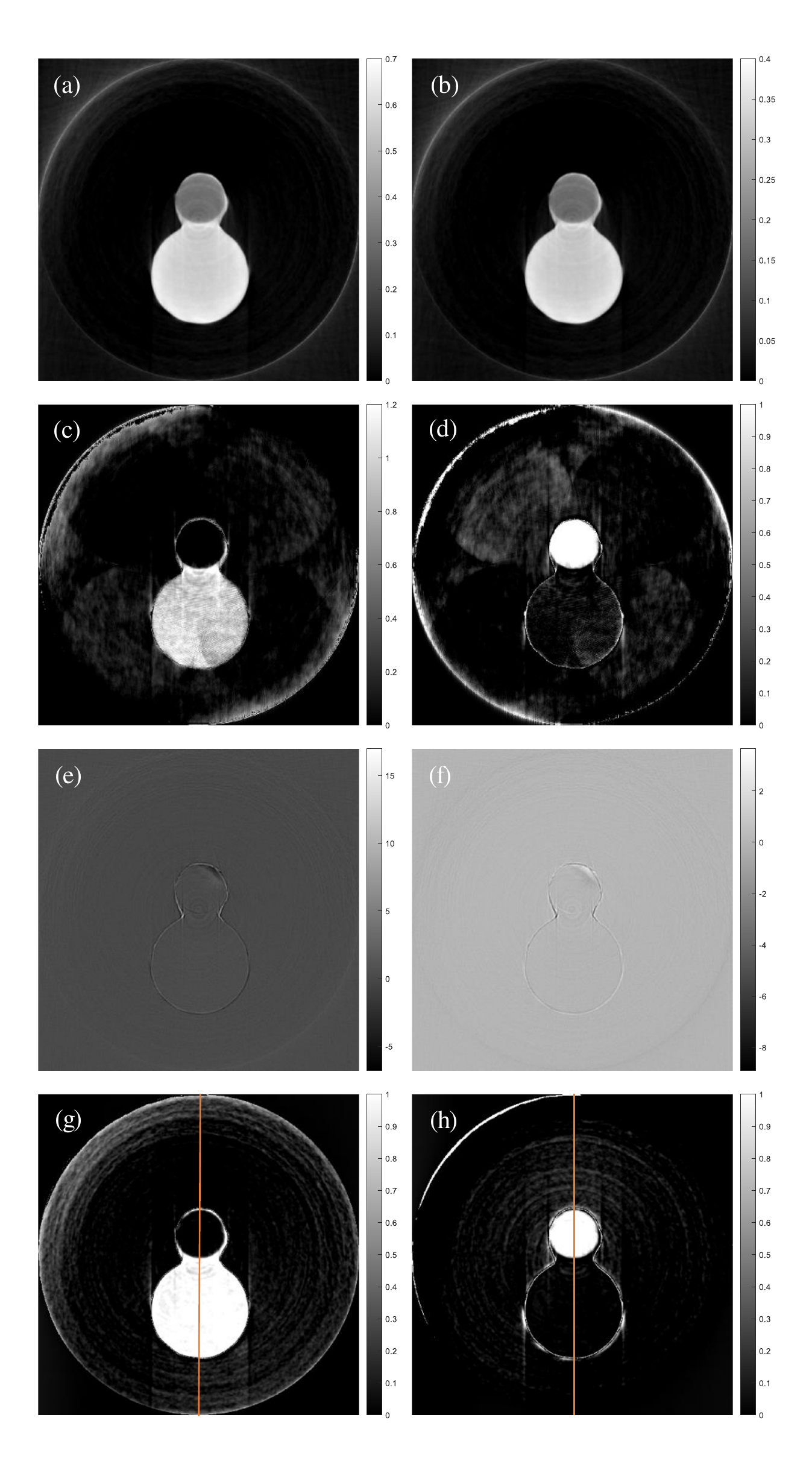}}
	\caption{Material decomposition results. (a), (c), (e) and (g) are the PMMA-based decomposed results of Born approximation with single ODD method, linear method, MD-SPBI and AR-PPCT, respectively; (b), (d), (f) and (h) are the LDPE-based decomposed results of Born approximation with single ODD method, linear method, MD-SPBI and AR-PPCT, respectively.}
	\label{Fig.13}
\end{figure}

\textcolor{black}{We extract one layer of the 3D phantom for the reconstruction of $\bm{\delta}$ and $\bm{\beta}$ images , and those retrieved projections are shown in Fig.\ref{Fig.11}. The reconstructed tomographies and material decomposition are shown in Fig.\ref{Fig.12} and Fig.\ref{Fig.13}, respectively.}

\begin{figure}[!t]
	\centerline{\includegraphics[width=\columnwidth]{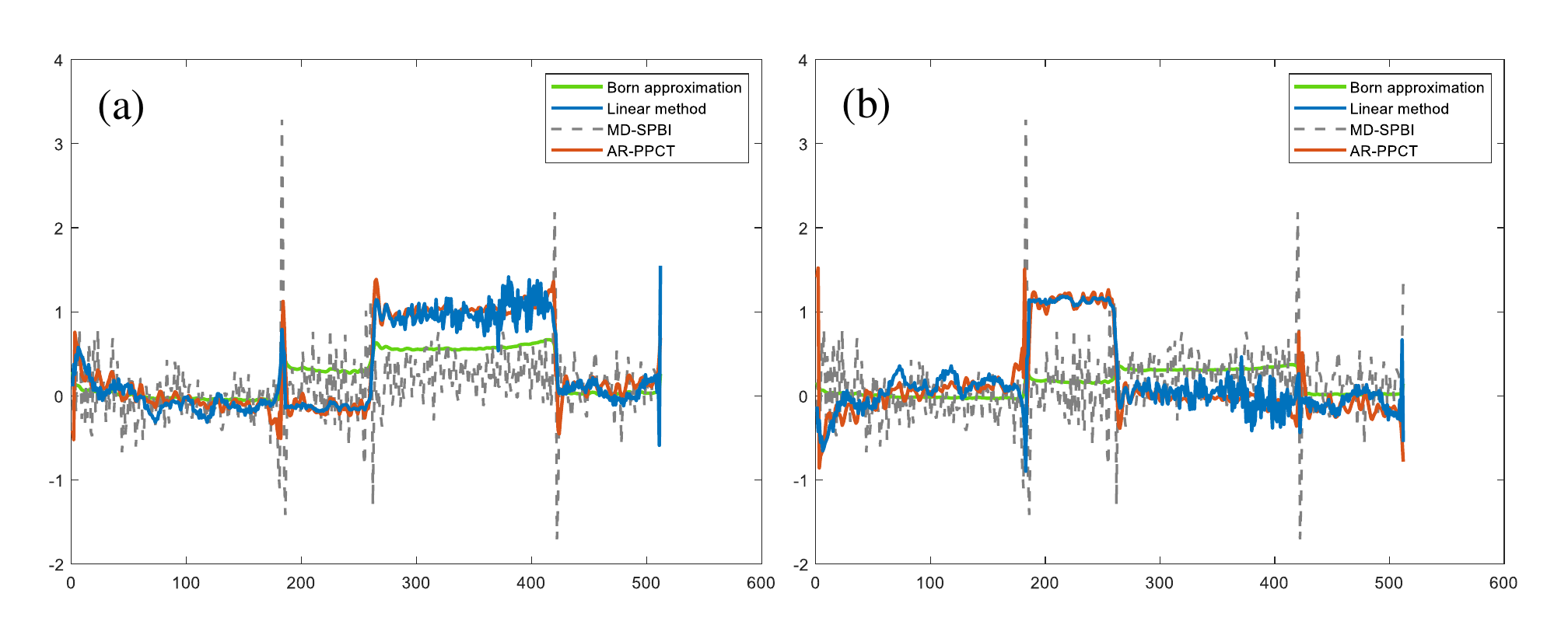}}
	\caption{Profiles of decomposed results. (a)The profiles of PMMA-based material results; (b)The profiles of LDPE-based material results. }
	\label{Fig.14}
\end{figure}

\textcolor{black}{Fig.\ref{Fig.12}(g-h) show the profiles of the $\bm{\beta}$ and $\bm{\delta}$ tomographies from orange lines in Fig.\ref{Fig.12}. \textcolor{black}{The material decomposition} \textcolor{black}{ results of the four methods are shown in Fig.\ref{Fig.13}.} \textcolor{black}{In Fig.\ref{Fig.14}, these are the profiles of the PMMA and LDPE images from the orange line in Fig.\ref{Fig.13}. 
Strong evidence of AR-PPCT‘s ability to decompose substances was found by comparing the profile results. There is a the relationship between $\mu$ and $\delta$, electron density ${\rho _{\rm{e}}}$ is the intermediate variable\cite{a55}:}}
\textcolor{black}{
\begin{equation}
\label{equ:decom2}
\left\{ {\begin{array}{*{20}{l}}
	{\mu (E,x,y) = {C_p}/{E^{{C_E}}}{\rho _e}(x,y)Z{{(x,y)}^{{C_z}}} + {C_{KN}}(E){\rho_e}(x,y)}\\
	{\delta (E,x,y) = {C_{PC}}(E){\rho_e}(x,y)}
	\end{array}} \right.,
\end{equation}
here $C_p$, $C_E$ and $C_z$ are the parameters to be determined. ${C_{PC}}(E) = \frac{{{r_0}{h^2}{c^2}}}{{2\pi {E^2}}}$, $r_0$ the classical radius of the electron, $h$ the Planck constant, $c$ the speed of light, and the Klein-Nishina cross section as follow,
\begin{equation}
\begin{split}
{C_{KN}}(E) = &2\pi r_0^2(\frac{{1 + a}}{{{a^2}}}(\frac{{2(1 + a)}}{{1 + 2a}} - \frac{{In(1 + 2a)}}{a}) + \frac{{In(1 + 2a)}}{{2a}}\\
 &- \frac{{1 + 3a}}{{{{(1 + 2a)}^2}}}),
\end{split}
\end{equation}
with $a = E/511keV$ the relative mass energy to electron. The theoretical equivalent atomic number Z for a compound was calculated by the following equation \cite{a56}:
\begin{equation}
\label{equ:decom3}
Z = {(\sum\limits_j {{w_j}Z_j^{2.94}} )^{1/2.94}},
\end{equation}
here $w_j$ the fraction of the total number of electrons associated with each element, and $Z_j$ the atomic number of element. The $\beta$ and $\delta$ coefficient of the two materials(PMMA, LDPE) are used in Eq.~(\ref{equ:decom2}) with 10 keV and 15 keV to fit the coefficient in the $\bm{\mu}$ formula. After fitting, the coefficients are $C_p=2.1086e^{-13}cm^2$, $C_E=3.337$ and $C_Z=3.673$.
As Table \ref{table.4} shows, compared with other methods, there is a significant that the fitted results of AR-PPCT are very close to the theoretical values.
The relative error is measured by the following formula:
\begin{equation}
    err = \frac{{\left| {{Z_T} - Z} \right|}}{{{Z_T}}}*100\% ,
\end{equation}
The relative errors of PMMA and LDPE in AR-PPCT are $0.9\%  \sim 2.7\% $  and $2.0\%  \sim 2.8\% $, respectively. In other words, the accuracy of basic materials is greater than $97.2\%$ in AR-PPCT.}

\begin{table}[]
	\caption{{Equivalent atomic number comparison of PMMA and LDPE materials}. ($Z_{T}$ is the theoretical value by (\ref{equ:decom3}).)}
	\centering
	\scalebox{0.65}{
		\begin{tabular}{cccccc}
			\toprule
			&$Z_{T}$&Born approximation&Linear method&MD-SPBI&AR-PPCT\\
			\cmidrule(r){2-6}
			\multirow{1}{*}{PMMA}&6.467&5.243$\pm0.07$&\textcolor{black}{6.397$\pm$0.35}&3.451$\pm$1.20& 6.523$\pm$0.12 \\
			\cmidrule(r){1-6}
			LDPE&5.444&2.510$\pm$0.03&\textcolor{black}{5.434$\pm$0.14}&3.057$\pm$0.95&5.424$\pm$0.13 \\
			\bottomrule
			\label{table.4}
	\end{tabular}}
\end{table}

\begin{figure}[!t]
	\centerline{\includegraphics[width=\columnwidth]{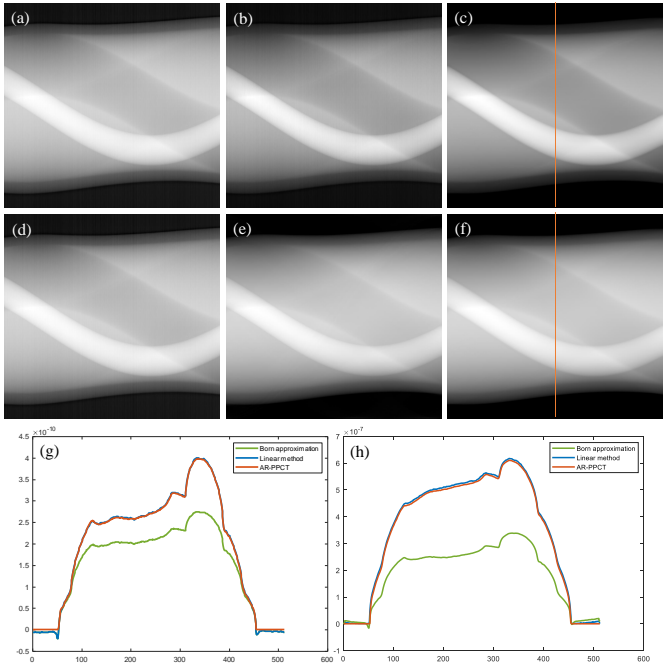}}
	\caption{Projections results. (a), (b) and (c) are the retrieved absorption maps of Born approximation with single ODD method, linear method and AR-PPCT, respectively; (d), (e) and (f) are the retrieved phase maps of Born approximation with single ODD method, linear method and AR-PPCT, respectively; (g) and (h) are the profile lines of absorption and phase maps, respectively.}
	\label{Fig.15}
\end{figure}

\subsection{Real experiment 2}

\textcolor{black}{The experiment were performed on the beamline BL13W at the Shanghai Synchrotron Radiation Facility (SSRF). The experimental parameters are displayed in Table \ref{table.30}. The data of 540 angles were collected at equal intervals within 180 degrees. The reconstructed image size was 512*512. The second energy was 12 keV in MD-SPBI.}

\begin{table}[]
	\caption{{The experimental parameters in the Shanghai Synchrotron Radiation Facility}(BL13W).}
	\centering
	\scalebox{0.8}{
		\begin{tabular}{cc}
			
			\toprule 
			
			\multirow{1}{*}{Energy}&20 keV\\
			
			\multirow{1}{*}{ODD}&35 cm\\
			
			\multirow{1}{*}{Pixel Array Detector}&2048*800\\
			
			\multirow{1}{*}{Detector Unit Size} &6.5 um \\
			
			\multirow{1}{*}{Single Exposure Time} &3.5 ms\\
			
			\multirow{1}{*}{Samples $\&$ Size} &PMMA: 5.6 mm; LDPE: 4.0 mm; \\&PTFE: 2.0 mm \\

			\bottomrule
			\label{table.30}
	\end{tabular}}
\end{table}

\begin{figure}[!t]
	\centerline{\includegraphics[width=\columnwidth]{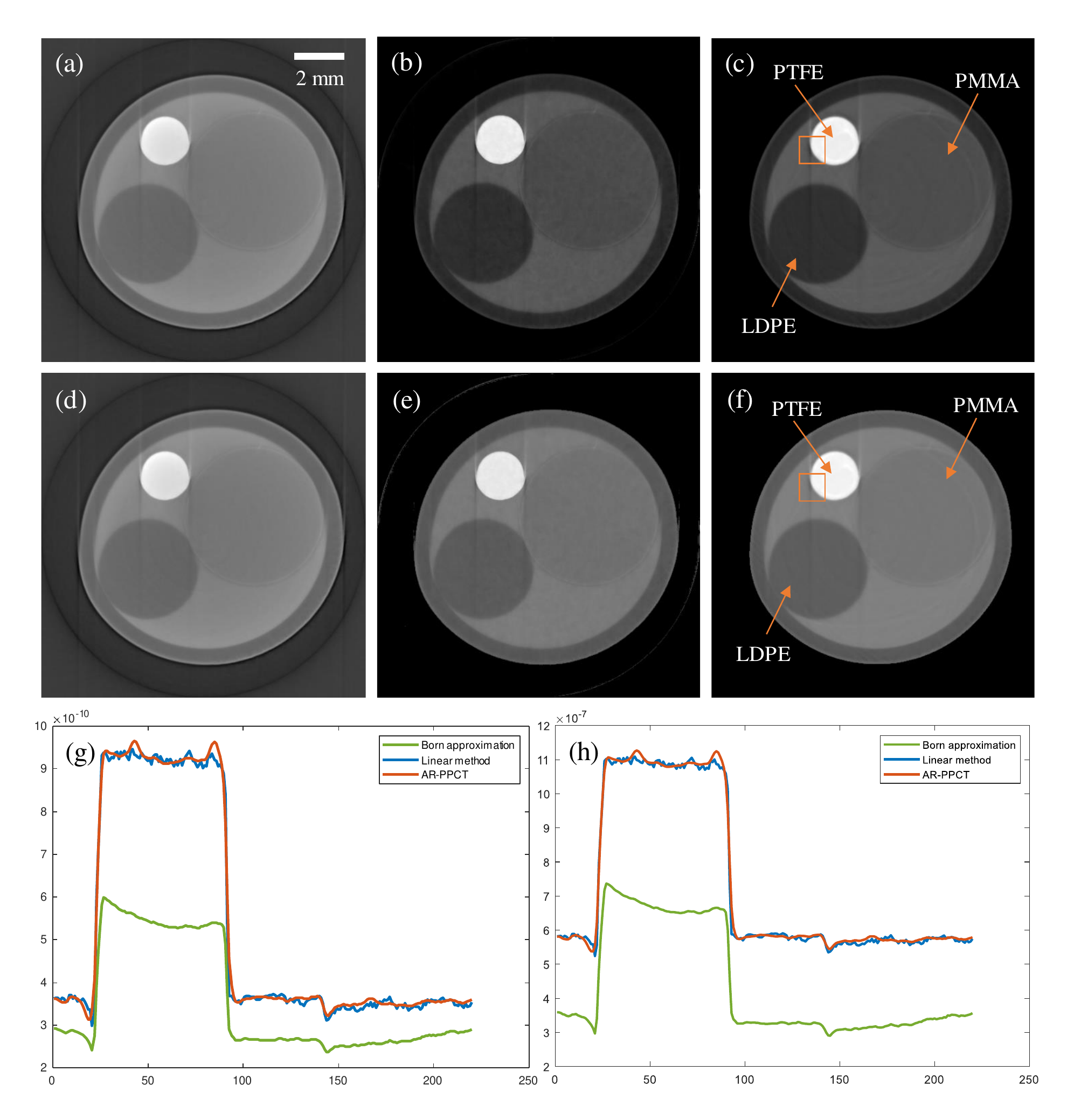}}
	\caption{Tomography results. (a)The $\bm{\beta}$ image of Born approximation with single ODD method; (b)The $\bm{\beta}$ image of Linear method; (c)The $\bm{\beta}$ image of AR-PPCT; (g)The profiles of $\bm{\beta}$ images; (d), (e) and (f) are the $\bm{\delta}$ image of Born approximation with single ODD method, Linear method and AR-PPCT, respectively; (h)The profiles of $\bm{\beta}$ images. }
	\label{Fig.16}
\end{figure}

\textcolor{black}{As shown in Fig. \ref{Fig.10}(c), the phantom consists of four components, LDPE, PMMA, PTFE and water. The PMMA, LDPE and PTFE cylinders with diameters of 5.6 mm, 4.0mm and 2.0 mm, respectively, were placed in a polyethylene plastic tube with an external diameter of 10.7 mm, then injected with water to form the whole sample. }

\begin{figure}[!t]
	\centerline{\includegraphics[width=\columnwidth]{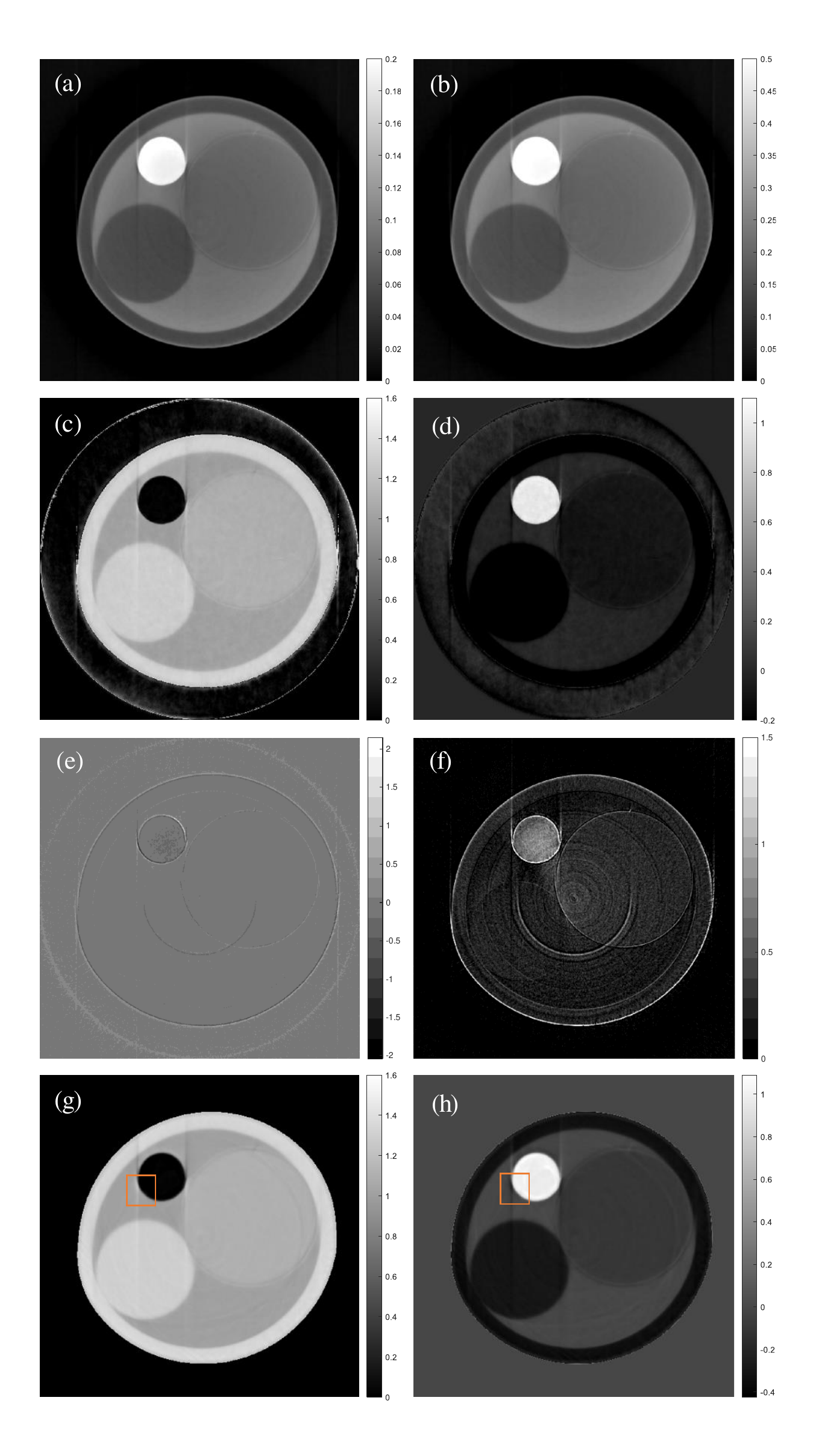}}
	\caption{Material decomposition results. (a), (c), (e) and (g) are the water-based decomposed results of born approximation method, linear method, MD-SPBI and AR-PPCT respectively; (b), (d), (f) and (h) are the PTFE-based decomposed results of born approximation method, linear method, MD-SPBI and AR-PPCT, respectively. }
	\label{Fig.17}
\end{figure}

\begin{figure}[!t]
	\centerline{\includegraphics[width=\columnwidth]{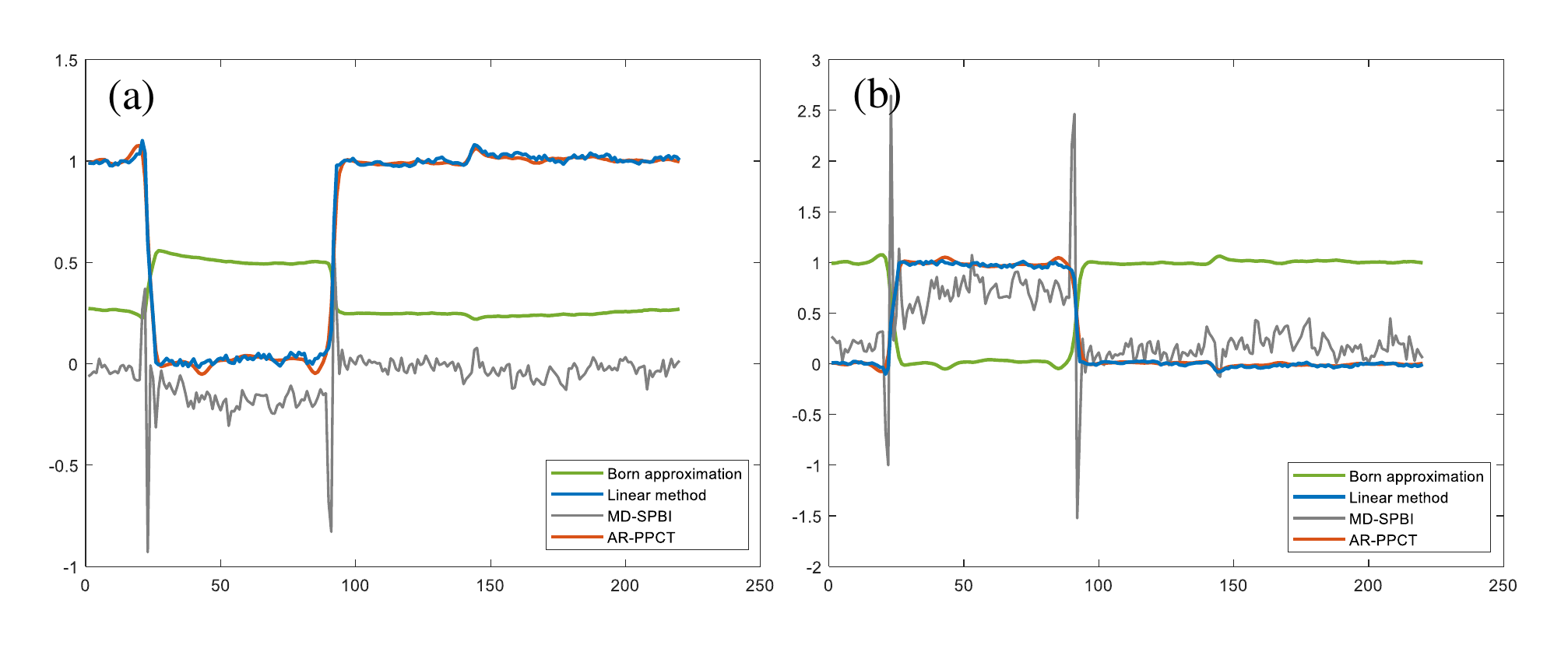}}
	\caption{Profiles of decomposed results. (a)The profiles of water-based material results; (b)The profiles of PTFE-based material results. }
	\label{Fig.18}
\end{figure}

\textcolor{black}{In this experiment, we used water and PTFE as substrates to perform AR-PPCT, MD-SPBI, Linear method and Born approximation with single ODD. Those retrieved projections are shown in Fig.\ref{Fig.15}. Fig. \ref{Fig.16} shows the $\bm{\beta}$  and  $\bm{\delta}$ images of those methods. The outer ring in the image represents the polyethylene plastic container, and there are three different components in the inner ring, namely, three circles with different gray levels in the middle of the ring, LDPE phantom with the lowest gray level on the left, PTFE phantom with the highest gray level at the top, PMMA phantom at the upper right and water at the rest. Fig.\ref{Fig.16}(g,h) show the profiles of the $\bm{\beta}$ and $\bm{\delta}$ tomographies from orange squares.  Fig.\ref{Fig.17} shows the decomposition results of different methods. Comparing the decomposition results of AR-PPCT, MD-SPBI, Linear method and Born approximation with single ODD methods, 
the most obvious finding was that AR-PPCT method has better material quantitation performance. As shown in Fig.\ref{Fig.18}, those are the profiles of the water and PTFE images from the orange squares in Fig.\ref{Fig.17}. These results provide important proof for the decomposition ability of materials in AR-PPCT.}

\section{Discussion} 
\textcolor{black}{In experiments, we compare AR-PPCT with three other quantitative methods. The comparison include the phase and \textcolor{black}{absorption projection map, tomographic results of $\bm{\beta}$ and $\bm{\delta}$, and base material results. It is remarkable that the results of material decomposition in AR-PPCT are better than those of the three comparative methods. Because there is a single fixed ratio between the information of phase shift and attenuation in Born approximation with single ODD, and it is hard to register two sets of data under different energies in MD-SPBI when the imaging system and the imaged object are mechanically unstable. \textcolor{black}{There are some artifacts and noise in the attenuation and material tomography in real experiments since the Linear method lacks feedback and correction.}
With calculating the equivalent atomic number $(Z)$, there is actually obvious difference between AR-PPCT and other methods. The results of AR-PPCT are very close to the theoretical value, by calculating the relative error, it can get the accuracy is greater than $97.2\%$. 
It is because the improved approximation $\delta=a*\beta+b$ plays a crucial role in this quantitation method.}}

\textcolor{black}{Unlike dual-energy CT and grating-based imaging,
AR-PPCT can effectively distinguish low-Z materials with one set of projection data. Besides, it is easy to implement since no additional requirement of optical components, such as gratings or special detectors (PCD) in the beam. The proposed method is better suitable for thin samples in this paper due to its reliance on Born approximation. However, it is the linear approximation of the real and imaginary components of the refractive index and the one-step concept that are not restricted by the sample size. They can be combined with other imaging methods that work well with thick samples to achieve high-quality quantitative imaging. Validation and comparison in polychromatic laboratory are beyond the scope of this study and will be investigated in the future}.

\textcolor{black}{The linear proportional relationship in AR-PPCT is a preliminary approximation, which cannot be accurately fitted to more than two base materials at a time. We also assume that $\delta  = {a_2}{\beta ^2} + {a_1}\beta  + {a_0}$, which means that the case of $N=2$ in Eq.~(\ref{equ:decom00}).  One of the obstacles is that it is hard to deal with the constant term $ - \frac{{2\pi }}{\lambda }\int {{a_0}} dl$. Even if the constant term is eliminated, the equation will also become difficult to solve for the residual term ${\mathop{\rm Re}\nolimits} {\psi ^\varphi } - {\mathop{\rm Re}\nolimits} {\psi ^{\varphi (m)}}$. Therefore, achieving higher precision multi-substrates phase retrieval is a potential research effort.}

\section{Conclusion} 
\textcolor{black}{In this paper, we propose a one-step method based on the Fresnel diffraction model that can reconstruct the images of the $\bm{\beta}$ and $\bm{\delta}$ simultaneously. Since the Fresnel propagator is a Gaussian-like function, the convolution operation indicates it plays a role in spreading and smoothing the wavefront in the evolution process.
With feedback and corrections in this iterative method, the noise can be effectively restrained. \textcolor{black}{Reconstruction algorithms based on optimization models impose some constraints on the image to reduce artifacts and noise. However, here we only focus on reconstruction based on the imaging model without constraints.}
Compared to multi-scanning methods, AR-PPCT definitely reduces the data acquisition time and processing difficulty.}
\textcolor{black}{Furthermore, it is a quantitative single-distance multi-materials reconstruction method. We utilize an improved additional approximation to obtain better quantitative reconstruction and decomposition results. Simulation and real experimental results verify that AR-PPCT outperforms the other quantitative methods, especially for the image quality and quantitative accuracy.}
\textcolor{black}{We anticipate that this algorithm has the potential for quantitative imaging research, especially for imaging live samples such as insects, mice, and human breast preclinical studies.}

\section*{Appendix}
\subsection{Derivation of the iterative algorithm} 
\textcolor{black}{For convenience, we analyze the situation of 2-dimensional samples here, and this method is also applicable to 3-dimensional samples. Utilizing the linear relation Eq.~(\ref{equ:decom0}), we can obtain:
\begin{equation}
    {\mathop{\rm Im}\nolimits} {\psi _0} = a{\mathop{\rm Re}\nolimits} {\psi _0} + \mathcal{F}\left[ { - k\int b dl} \right],
\end{equation}
Substituting the above formula into Eq.~(\ref{equ:de2})
\begin{equation}
\begin{split}
    \mathcal{F}\left[ {\frac{{{I_z}}}{{{I_0}}}{\rm{ - 1}}} \right] = 2\cos (z{\rho ^2}){\mathop{\rm Re}\nolimits} {\psi _0} + 2a*\sin (z{\rho ^2}){\mathop{\rm Re}\nolimits} {\psi _0} \\
    + {\rm{2}}\sin (z{\rho ^2})\mathcal{F}\left[ { - k\int b dl} \right],
  \end{split}
\end{equation}}

\textcolor{black}{The intensity residual of the m-th iteration at projection angle $\varphi$ is:
\begin{equation}
    {\mathop{\rm Re}\nolimits} _z^{\varphi (m)} = I_z^\varphi  - {\left| {{h_z} \otimes \left[ {{A^{in}}\exp ( - \frac{{{M^{\varphi (m)}}}}{2} + {\bf{i}}{\Phi ^{\varphi (m)}})} \right]} \right|^2},
\end{equation}
\begin{equation}
    \begin{array}{l}
     \mathcal{F}({\mathop{\rm Re}\nolimits} _z^{\varphi (m)}) = \mathcal{F}\left( {I_z^\varphi  - I_z^{\varphi (m)}} \right)\\
     {\rm{               }} = \left[ {2\cos (z{\rho ^2}) + 2a\sin (z{\rho ^2})} \right]\left( {{\mathop{\rm Re}\nolimits} {\psi ^\varphi } - {{{\mathop{\rm Re}\nolimits} }^{\varphi (m)}}} \right),
\end{array}
\end{equation}
The absorption residual of the m-th iteration at projection angle $\varphi$ is:
\begin{equation}
    {\mathop{\rm Re}\nolimits} {\psi ^\varphi } - {{\mathop{\rm Re}\nolimits} ^{\varphi (m)}} = \frac{{\mathcal{F}({\mathop{\rm Re}\nolimits} _z^{\varphi (m)}/2)}}{{\cos (z{\rho ^2}) + \sin (z{\rho ^2})*a}},
\end{equation}
Then we combine the residual with SART to get its iterative format.
}

\section*{Acknowledgements}
\textcolor{black}{This work was partially carried out with the support of Shanghai Synchrotron Radiation Facility. The authors thank Prof. Dr. Biao Deng for the help and support during the experiments at SSRF.}

\end{document}